\pdfoutput=1
\documentclass[conference,onecolumn]{IEEEtran}
\usepackage{graphicx}
\usepackage{multirow}
\usepackage{array}
\usepackage{enumerate}
\usepackage{diagbox}

\usepackage[font=bf]{caption}
\usepackage{cite}
\usepackage{url}
\usepackage{amsmath}
\usepackage{afterpage}

\usepackage{mathtools}
\DeclarePairedDelimiter{\floor}{\lfloor}{\rfloor}

\usepackage[caption=false,font=footnotesize,labelformat=simple]{subfig}
  
\usepackage{color}
\usepackage[table]{xcolor}

\begin{document}
\title{Unlocking the Potential of QoS-Aware Pricing under the Licensed Shared Access Regime}

\author{\IEEEauthorblockN{Vaggelis G. Douros, Andra M. Voicu, Petri M\"{a}h\"{o}nen\\}
\IEEEauthorblockA{Institute for Networked Systems, RWTH Aachen University\\
Kackertstrasse 9, 52072 Aachen, 
Germany\\
E-mail: \{vaggelis.douros, avo, pma\}@inets.rwth-aachen.de}}

% make the title area
\maketitle 

\begin{abstract}
We present a techno-economic analysis of a cellular market that operates under the licensed shared access (LSA) regime, consisting of a mobile network operator (MNO) that leases spectrum to a number of Programme Making and Special Events (PMSE) users. The MNO offers two \mbox{quality-of-service} (QoS) classes (high and low), differentiating the price based on the QoS class. The key question that we address is whether and to which extent the MNO has incentive to adopt this form of QoS-aware pricing. The first step is to model the parameters that are controlled by each PMSE user: i)~the way to choose between the two QoS classes and ii)~the available budget per QoS class. The second step is to compute the maximum revenue of the MNO. Our analysis reveals that the MNO can always tune the prices so as to maximise its revenue for the scenario where all users belong to the high QoS class. This is a consistent result throughout our study, that holds for any considered set of user-controlled parameters and of technical parameters. We conclude that the adoption of QoS-aware pricing in the LSA market generates a tussle between the MNO and the regulator. The MNO has incentive to support fewer users but with high QoS and charge them more, which is not aligned with the regulator's goal for social welfare maximisation. 
\end{abstract}
\IEEEpeerreviewmaketitle

\section{Introduction and Related Work}
Licensed shared access (LSA)~\cite{ECC2014} has been adopted in Europe as a promising paradigm to dynamically share licensed spectrum between different networks and technologies.  
LSA proposes a two-tier approach where the initial target use case considered mobile network  operators (MNOs) leasing spectrum in the 2.3--2.4~GHz band from incumbent technologies like Programme Making and Special Events (PMSE)~\cite{ECC2015}. 
However, recent initiatives from industry and spectrum regulators have proposed a symmetric use case, where PMSE users could lease spectrum from MNOs, targeting reliable short-term use of spectrum for concerts, conferences, etc.~\cite{PMSExG2017}. 
 
Though the adoption of LSA brings significant benefits from a technical perspective, a number of business challenges arise for the key stakeholders of the market (\emph{i.e.}, regulator, incumbent spectrum user, and LSA licensee). These include the MNO's costs of additional infrastructure and the required modifications of the existing systems to support and manage the sharing procedure, as well as the license fees~\cite{tehrani2016}. Thus, the stakeholders must perform a \mbox{techno-economic} analysis in order to assess whether LSA is worth the investment. However, business research on LSA is scarce~\cite{ahokangas2014,ahokangas2014b,matinmikko2018} and focuses on the qualitative domain, without offering quantitative results on whether LSA schemes are \mbox{techno-economically attractive.} 

The work closest to ours is~\cite{voicu2018}, where an MNO that operates under the LSA framework leases spectrum to a number of PMSE users that belong to two distinct quality-of-service (QoS) classes, admitting either low or high QoS requirements. As in~\cite{voicu2018}, we study scenarios where all users have either high or low QoS requirements, as well as mixed QoS requirements (\emph{i.e.}, some users have low and some users have high QoS requirements).  We extend the approach of~\cite{voicu2018}, aiming at unlocking the potential of QoS-aware pricing in this LSA market, where we adopt price differentiation based on the QoS class. Our key contributions are the following. From the perspective of the PMSE users, we model the behaviour of the users regarding how they choose between the two QoS classes, as well as their available budgets for the two QoS classes. Through this process, we are able to predict the distribution of the users between the two QoS classes for each possible combination of \mbox{considered prices.}  

From the perspective of the MNO, we identify the prices that correspond to the maximum revenue that can be achieved for each QoS scenario. A consistent result arises independently of i)~the distribution of the budgets, ii)~the way that the users choose between the QoS classes, and iii)~the values of the technical parameters. The MNO can always tune the prices so that the maximum revenue for the high QoS scenario is the highest, followed by the mixed QoS scenario and finally by the low QoS scenario. This result highlights the potential of \mbox{QoS-aware} pricing for the MNO, since the MNO has motivation to sacrifice some of the users with low QoS in order to support more users with high QoS and charge them more. This is also interesting from a regulatory point of view, since we identify a constant tussle in the LSA market, where the goal of the MNO (\emph{i.e.}, revenue maximisation) is not aligned with the goal of the market regulator (\emph{i.e.}, social welfare maximisation). Finally, we quantify the impact of the budget parameters on the revenue of the QoS scenarios, providing insights for which markets have the potential to be more profitable for the MNO. 

\section{The Techno-Economic Problem}

We first summarise the techno-economic input from~\cite{voicu2018} that we are going to use for our analysis. Then, we introduce our extensions. We assume a monopolistic market with one MNO and $N$ PMSE users that are interested in leasing spectrum from the unique MNO. Consistent with one of the business models in~\cite{PMSExG2017}, the PMSE users also utilise the network infrastructure of the MNOs. Furthermore, the PMSE users are classified into two distinct QoS classes: there are at most \mbox{$N_L$ PMSE users} with low QoS requirements (e.g., audio speech applications) and at most $N_H$ PMSE users with high QoS requirements (e.g., high definition audio productions). We are interested in analysing from a techno-economic point of view the following three \mbox{QoS scenarios:}
\begin{itemize}
\item \emph{Low QoS Scenario}: The MNO can support at most $N_L$ users, where all of them have the same low QoS requirements $Q_L$.
\item \emph{High QoS Scenario}: The MNO can support at most $N_H$  users, where all of them have the same high QoS requirements $Q_H$.
\item \emph{Mixed QoS Scenario}: The MNO supports users with mixed QoS requirements, \emph{i.e.}, at most $N_\mathit{L,M}$ users with $Q_L$ and at most $N_\mathit{H,M}$ users \mbox{with $Q_H$.}
\end{itemize}
Given the maximum number of supported PMSE users for the three QoS scenarios, the goal of the MNO is to define a pricing policy and choose the scenario that will maximise its revenue. Among the four pricing policies that have been considered in~\cite{voicu2018}, we apply QoS-aware pricing, where the differentiation in the price is based on the QoS class that each user belongs to \cite{huang2013}. Depending on the assumptions and the model, QoS-aware pricing may maximise e.g. the revenue of the MNO or the social welfare \cite{shetty2010, wang2017}. 

We adopt a type of QoS-aware pricing which corresponds to an application of the \emph{second degree of price discrimination}~\cite{maille2014}. In this form of discrimination, there are at least two distinct prices, which correspond to at least two different types of services. Any customer who wants the same type of service will pay the same price. In our case, we propose that the discrimination is based on the QoS class that each PMSE user belongs to; each user that targets $Q_L$ pays $P_L \in [P_{L,\min},P_{L,\max}]$, whereas each user that targets $Q_H$ pays $P_H$. We also define parameter $K=\frac{P_H}{P_L}$ which is always above 1. Then, the revenue of the MNO for each of the three QoS scenarios is:
\begin{align}
\text{Low QoS Scenario: }& N_LP_L, \label{eq:lowQoS} \\
\text{High QoS Scenario: }& N_HP_H=N_HKP_L,\label{eq:highQoS} \\
\text{Mixed QoS Scenario: }& N_\mathit{L,M}P_L+N_\mathit{H,M}P_H=N_\mathit{L,M}P_L+N_\mathit{H,M}KP_L. \label{eq:mixedQoS} 
%\vspace{-2cm}
\end{align}
Clearly, the scenario that maximises the MNO's revenue can be computed by the following formula:
 \begin{equation*} \max\{N_L, N_HK, N_\mathit{L,M}+N_\mathit{H,M}K\}.\end{equation*}

In~\cite{voicu2018}, there has been an extensive study of the revenue for the three QoS scenarios. For different values of the technical parameters including carrier frequency $f$, propagation environment, base station (BS) transmit power level, and bandwidth, the maximum number of supported PMSE users for the three QoS scenarios has been computed. Then, the revenue after the application of QoS-aware pricing has been estimated for a fixed value of $P_L$  and a range of values of $P_H$. A key assumption during the whole analysis was that the MNO always serves the maximum number of users that can be \mbox{technically supported.}%(\emph{i.e.}, for different $K$). 

We generalise this study towards the following two directions. First, we introduce an additional degree of freedom studying markets with different values of $P_L$. Second, we relax the assumption that the market always performs at its maximum capacity by proposing a methodology to compute the exact number of PMSE users that will be admitted in each QoS scenario. In order to do so, we need to model the behaviour of the users. Initially, we need to model how a user chooses between the two QoS classes. Therefore, we introduce a metric $w$ that quantifies the preference of each user $i$ for each QoS class by weighing the importance that the user gives to the price and the QoS. 
For the high QoS class, $w$ is defined as follows: 
\begin{equation*}
w_{H,i}=a_i\frac{P_L}{P_L+P_H}+(1-a_{i})\frac{Q_H}{Q_L+Q_H},
\end{equation*} 
where the user-specific parameter $a_i $ follows a uniform distribution in (0,1). When $a_i$ is above 0.5, user $i$ considers as the most important factor the price that it has to pay, otherwise the most decisive factor is the QoS that it gets. We note that we use fractions for a relative comparison of the two factors that influence the decision of the user, which is why $w$ also ranges between 0 \mbox{and 1.} 

Similarly, for the low QoS class,  $w$ is defined as:
\begin{equation*}
w_{L,i}=a_i\frac{P_H}{P_L+P_H}+(1-a_i)\frac{Q_L}{Q_L+Q_H}.
\end{equation*} 
Note that $w_{H,i}+w_{L,i}=1$, meaning that each user $i$ needs to compute just one of them. If $w_{H,i}$ is higher than 0.5, then user $i$ prefers the high QoS class. Otherwise, it prefers the low QoS class.

Another aspect that was not modelled in~\cite{voicu2018} is the user's available budget for each QoS class. Though we are not aware of specific studies for the distribution of the budgets of the PMSE users, we expect that it follows a (variation of the) normal distribution. This is in accordance with adjacent telecommunication markets~\cite{maille2014}. More specifically, we model the distribution of the budget for the low QoS $B_L$ as a truncated normal distribution with minimum value \mbox{$P_{L,\min}=\$10$ ~\cite{Ofcom2018}.}  We need a minimum value, otherwise a user can never get access to this QoS class, so it is not of interest for this market. We study 6 cases for $B_L$, where the mean $\mu_L=\{0.5,0.7,0.9\}P_{L,\max}$ and the standard deviation $\sigma_L=\{0.2,0.4\}P_{L,\max}$, with $P_{L,\max}=\$120$~\cite{Ofcom2018}. 

Then, we model the distribution of the budget for the high QoS $B_H$ as a truncated normal distribution with minimum value $B_{L}$. The motivation for this minimum threshold is that the user's budget for the high QoS class should be at least equal to its budget for the low QoS class. For $B_H$, we also consider 6 cases, where the mean $\mu_H=\{0.2,0.4,0.6\}\frac{Q_H}{Q_L}B_{L}$ and the standard deviation $\sigma_H=\{0.2,0.4\}\frac{Q_H}{Q_L}B_{L}$. The quantity $\frac{Q_H}{Q_L}B_{L}$ is used as a benchmark, since, as we know from adjacent markets \cite{maille2014}, a typical user is expected to be willing to spend at most $\frac{Q_H}{Q_L}$ times more to get the class $Q_H$ instead of the class $Q_L$. Moreover, since the budget of the users for more expensive services is expected to be tighter, the coefficients of $\mu_H$ are typically lower than the ones of $\mu_L$.

\subsection*{Maximum Number of PMSE Users}
Table~\ref{table_qos} summarises the values of the technical parameters from \cite{voicu2018} used to estimate the maximum number of PMSE users that can be \mbox{technically supported.} 
Each PMSE user has either high or low QoS requirements. We define the QoS requirements in terms of the target Application-layer throughput $R$, where high QoS and low QoS correspond to 4.61~Mbps and 150~kbps, respectively. These values are consistent with the highest and lowest PMSE audio throughput requirements in~\cite{3GPP2018, PMSExG2017}, where low throughput values correspond to audio speech applications, while high throughput values are required for high definition audio productions~\cite{Pilz2018}.  Based on these values of the technical parameters, Table~\ref{table_max_users} summarises from \cite{voicu2018} the maximum number of users that can be supported for the three QoS scenarios. Since the number of users for the carrier frequencies of 2600~MHz and 3800~MHz are quite similar, we analyse only three cases: i)~800~MHz for the indoor propagation environment, ii)~800~MHz for the outdoor propagation environment, and iii)~3800~MHz for the indoor \mbox{propagation environment.}

\begin{table}[t]
\caption{PMSE user QoS requirements and technical parameters.}
\label{table_qos}
\centering
\begin{tabular}{|p{4cm}|p{2.2cm}|p{2.3cm}|p{3.5cm}|}
  \hline
  \multirow{2}{*}{\parbox{4cm}{\centering\textbf{Parameter}}} & \multicolumn{3}{c|}{\textbf{Value}} \\
  \cline{2-4}
   & \parbox[c][0.9cm]{2.2cm}{\centering\textbf{Low QoS Scenario}} & \parbox{2.2cm}{\centering\textbf{High QoS Scenario}} & \parbox{3.8cm}{\centering\textbf{Mixed QoS \\Scenario}} \\
  \hline
  PMSE user QoS requirements as Application-layer throughput $R$ & 150~kbps \cite{3GPP2018, PMSExG2017} & 4.61~Mbps \cite{3GPP2018, PMSExG2017} & 4.61 Mbps for 50\% of the users in the high QoS scenario and 150~kbps for other users\\
  \hline
  bandwidth $C$ &  \multicolumn{3}{l|}{20 MHz~\cite{3GPP2017}}\\
  \hline
  carrier frequency $f$ & \multicolumn{3}{l|}{800, 2600, 3800 MHz~\cite{3GPP2017}}\\
  \hline
  BS transmit power  $T$ &  \multicolumn{3}{l|}{30 dBm~\cite{3GPP2017, 3GPP2017a} (same for all BSs)}\\
  \hline
  propagation environment & \multicolumn{3}{l|}{indoor, outdoor} \\
 \hline
\end{tabular}
%\vspace{-0.3cm}
\end{table} 

\begin{table} [t]
\caption{Max. number of users that can be supported for the three QoS scenarios for the different values of the technical parameters.}
\label{table_max_users}
\begin{center}
\footnotesize
\begin{tabular}{ |p{4.3cm} || c|| c || c | c |}  
 \hline
  \multirow{3}{*}{\diagbox[width=4.74cm, height=1cm]{\textbf{Frequency,}\\ \textbf{Environment}}{\textbf{Scenario}}} & \textbf{Low QoS}  & \textbf{High QoS} & \multicolumn{2}{c|}{\textbf{Mixed QoS}}\\
  & & & \multicolumn{2}{c|}{ } \\
\cline{4-5} 
  & \textbf{Users} $N_L$ & \textbf{Users} $N_H$ & \textbf{Users} $N_\mathit{L,M}$ & \textbf{Users} $N_\mathit{H,M}$ \\
\hline 
 {\cellcolor{green!25}} $f$=800~MHz, indoor & {\cellcolor{green!25}}65 & {\cellcolor{green!25}}6 & {\cellcolor{green!25}}21 & {\cellcolor{green!25}}3 \\ 
\hline
 {\cellcolor{green!25}} $f$=800~MHz, outdoor & {\cellcolor{green!25}}7 & {\cellcolor{green!25}}2 & {\cellcolor{green!25}}4 & {\cellcolor{green!25}}1 \\ 
\hline
 $f$=2600~MHz, indoor & 36 &   4 &  13 & 2 \\
\hline
 $f$=2600~MHz, outdoor & 31 &  4 &  12 &  2 \\
\hline
 {\cellcolor{green!25}} $f$=3800~MHz, indoor & {\cellcolor{green!25}}37 & {\cellcolor{green!25}}4 & {\cellcolor{green!25}}13 & {\cellcolor{green!25}}2 \\ 
\hline
 $f$=3800~MHz, outdoor & 33 & 4 &   12 &  2 \\
\hline
\end{tabular}
\end{center} 
\vspace{-0.8cm}
\end{table} 

\section{Revenue Analysis: A Case Study}
In this section, we illustrate the evolution of the revenue for the three QoS scenarios for the example of the carrier frequency $f$=3800~MHz and the indoor propagation environment. We assume that the market consists of 41 PMSE users so that, provided that all of them have the necessary budget to pay for the prices $P_L$ and $P_H$, the maximum number of supported users can be admitted (\emph{i.e.}, either $N_L$=37, or $N_H$=4). For a given set of prices $P_L$ and $P_H$, we assume that the users follow a so-called \emph{non-strict} version for the choice of the QoS class. In this non-strict version, a user initially applies for getting access to the QoS class that it prefers more based on the value of the weighted metric $w$. It gets access to this QoS class provided that the following two conditions hold: i)~it can afford to pay  the price that the MNO has announced and ii)~the MNO has not reached the maximum number of PMSE users that it can support for this QoS class. If the user does not get access to the QoS class of its first choice, then it applies for the other QoS class and it gets admitted provided that the same conditions hold. In the following section, we also consider a \emph{strict} version for the choice of the QoS class, where each user applies for only one QoS class, \emph{i.e.}, the one that corresponds to the highest value of the weighted metric $w$.

\begin{figure}[t!]
\centering
\subfloat[$P_L$=\$30]{\includegraphics[width=0.4\textwidth]{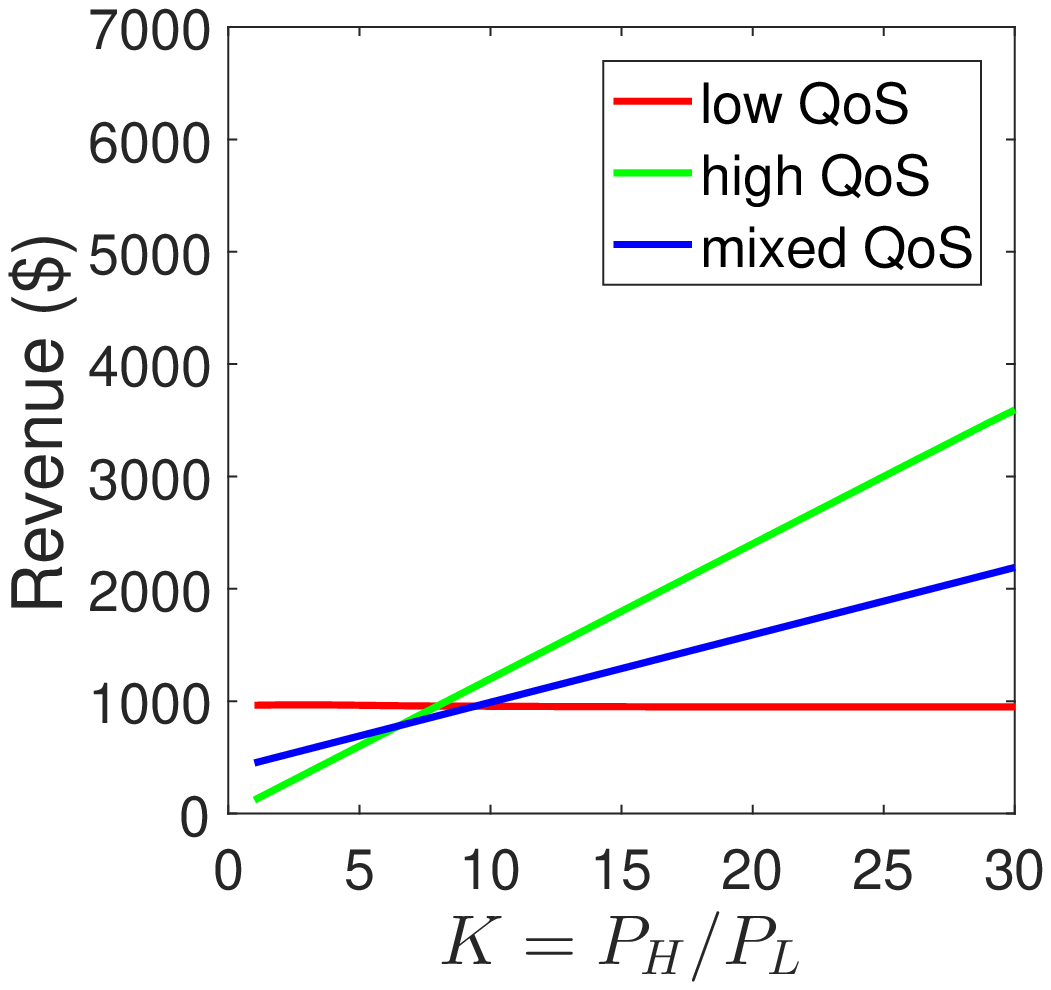} \label{fig_example_PL30}}
~
\subfloat[$P_L$=\$60]{\includegraphics[width=0.4\textwidth]{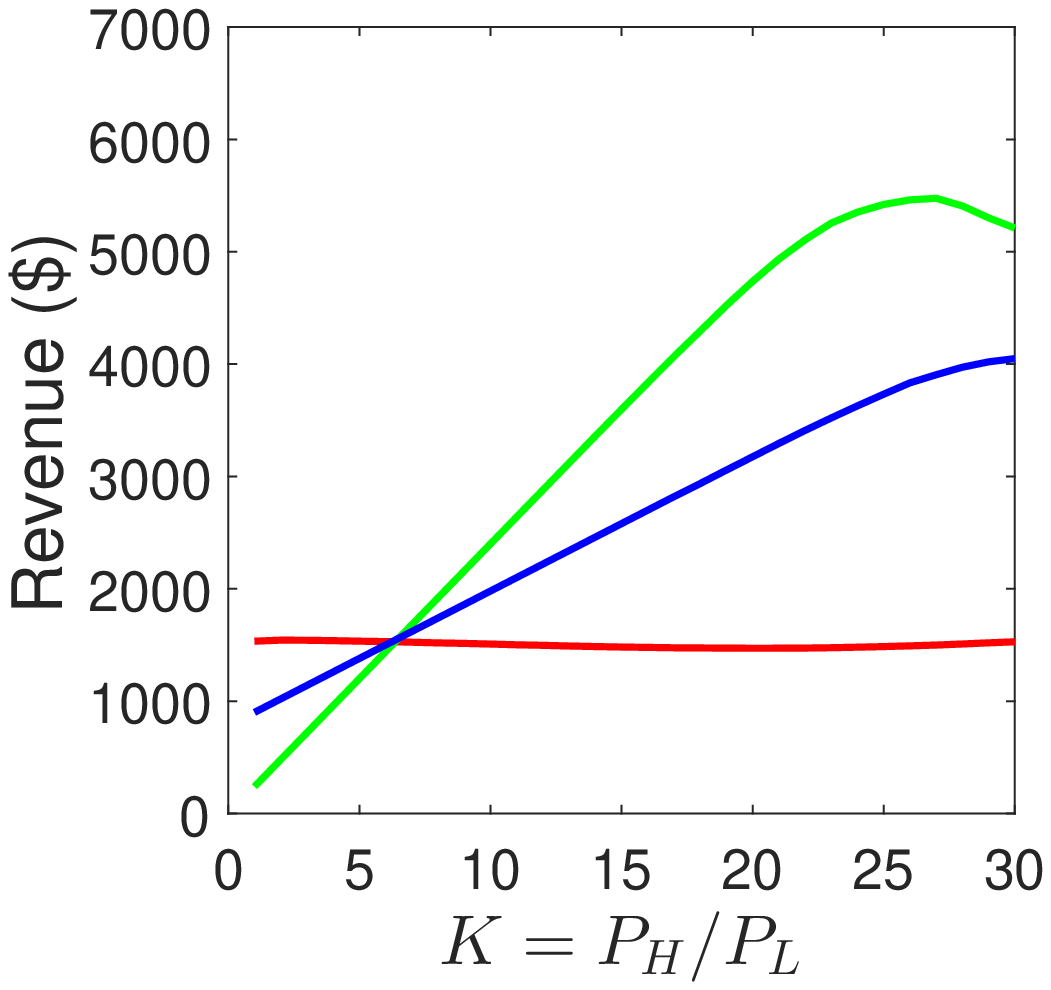} \label{fig_example_PL60}}\\
\subfloat[$P_L$=\$90]{\includegraphics[width=0.4\textwidth]{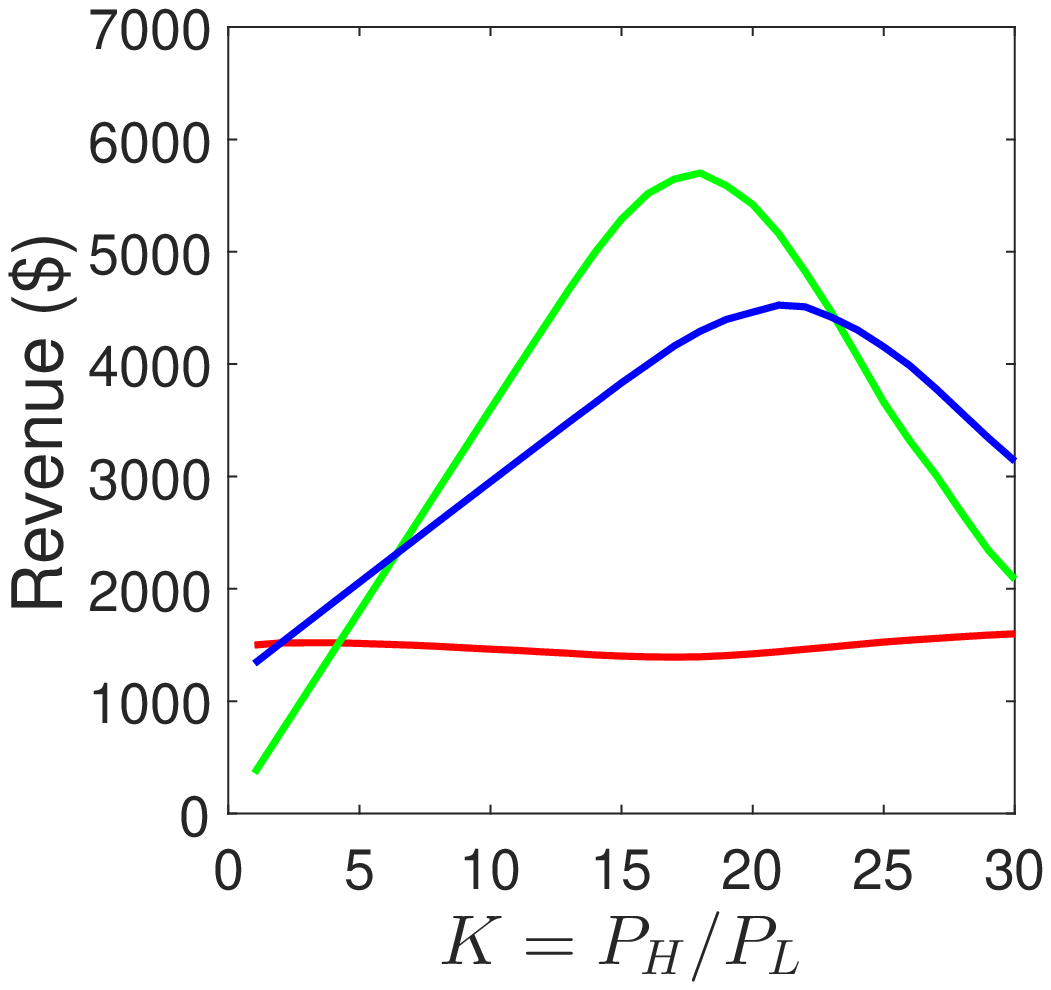} \label{fig_example_PL90}}
~\subfloat[$P_L$=\$120]{\includegraphics[width=0.4\textwidth]{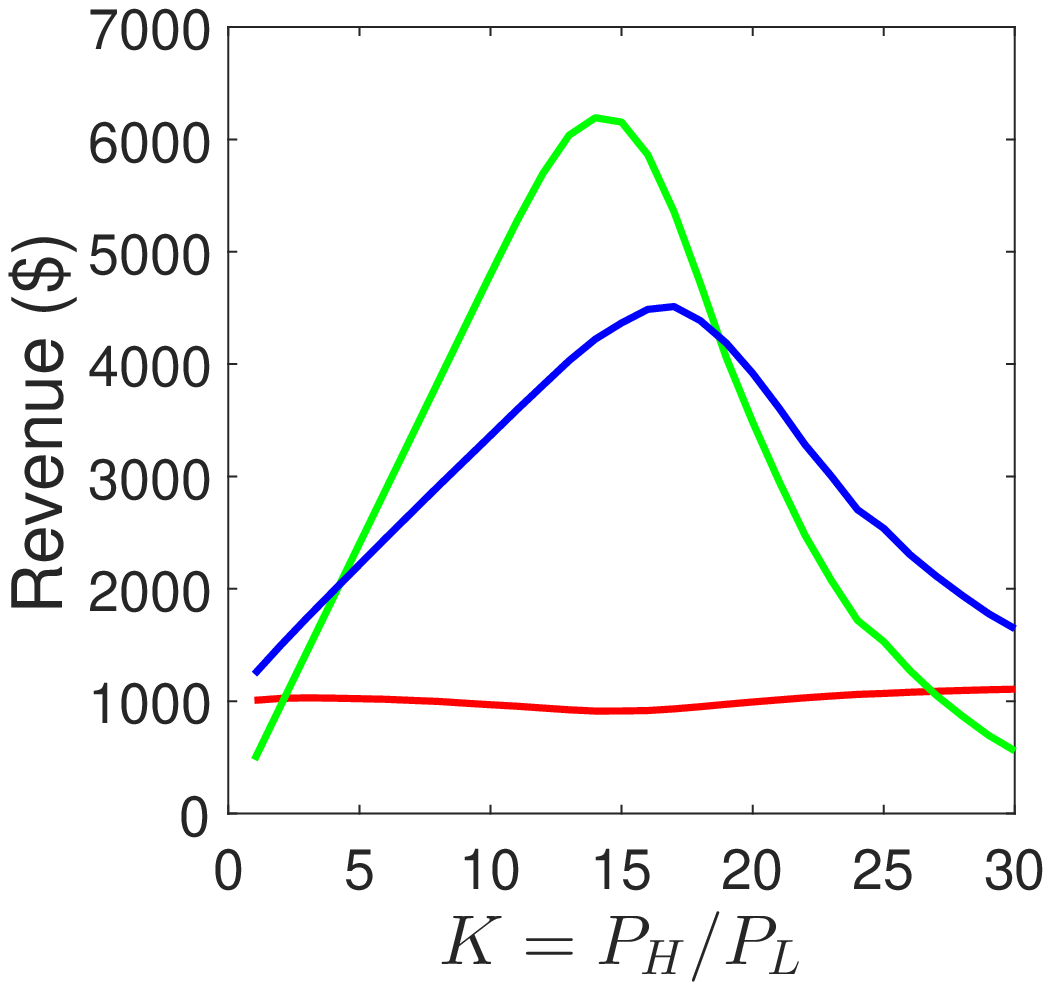} \label{fig_example_PL120}}
\\\caption{Evolution of the revenue for the three QoS scenarios. Technical parameters: carrier frequency $f$=3800~MHz and indoor propagation environment. Parameters of the distribution of the budgets $B_L$ and $B_H$: $\mu_L=0.7P_{L,\max}$,  $\sigma_L=0.4P_{L,\max}$, $\mu_H=0.4\frac{Q_H}{Q_L}B_{L}$,  $\sigma_H=0.2\frac{Q_H}{Q_L}B_{L}$. The choice of the QoS class is non-strict.}
\label{fig_revenue_example}
%\vspace{-3cm}
\end{figure}   

After deciding whether a user will be admitted and, if so, in which QoS class, the MNO computes the revenue for the three QoS scenarios. We consider four values of $P_L$, corresponding to 30, 60, 90, and 120 \$ for 48-hour access \cite{Ofcom2018}. For a given $P_L$, we apply QoS-aware pricing where $P_H=KP_L$, with parameter $K \in \{2, 3, \dots, \floor[\Big]{\frac{Q_H}{Q_L}}=30\}$.

Fig.~\ref{fig_revenue_example} shows the evolution of the revenue for the three QoS scenarios for the four values of $P_L$. Each subfigure corresponds to the revenue as a function of parameter $K$, for a given $P_L$. The results are averaged based on the simulation of 1000 markets, each consisting of 41 users. As we notice from Fig.~\ref{fig_example_PL30}, when parameter $K$ is below~7, the low QoS scenario generates the highest revenue. This is justified since the price differentiation between $Q_H$ and $Q_L$ is small enough to not overcome the difference between the actual number of users that are supported for $Q_H$ and $Q_L$. For higher values of $K$, the high QoS scenario generates the highest revenue, followed by the mixed QoS scenario. Also, the revenue for both the high QoS and the mixed QoS scenario increases linearly with $K$. This is expected from the corresponding equations \eqref{eq:highQoS} and \eqref{eq:mixedQoS} provided that the number of users $N_H$ and $N_{H,M}$ does not change with $K$. Finally, for the low QoS scenario, the revenue does not change with $K$, so any fluctuation is due to changes in the number of users. 

Fig.~\ref{fig_example_PL60} shows the revenue for $P_L=\$60$, where we notice some differences in the trends. First, though $P_L$ was doubled compared to Fig.~\ref{fig_example_PL30}, the revenue for the low QoS scenario was not doubled. This means that the budget $B_L$ of some of the users is below~\$60 and, therefore, they cannot afford to pay for this QoS class. Due to this, the high QoS scenario generates the highest revenue starting with a smaller value of $K$ (it is for $K>6$, whereas for $P_L=\$30$ it was for $K>7$). Moreover, for high values of $K$, the revenue for the high QoS scenario starts increasing sub-linearly and then it decreases. This is again due to budget constraints, this time for the budget $B_H$. The trend of a sub-linear increase is also noticed for the mixed QoS scenario, though it starts for higher values of $K$ compared to the high QoS scenario. This is expected since, for the mixed QoS scenario, the maximum number of users with high QoS that can be admitted is 2 instead of~4 for the high QoS scenario (see Table~\ref{table_max_users}). Therefore, for higher values of $K$, it is easier to find 2 instead of 4 users with $Q_H$. 

\begin{figure}
\centering
\subfloat[Budget low]{\includegraphics[width=0.4\textwidth]{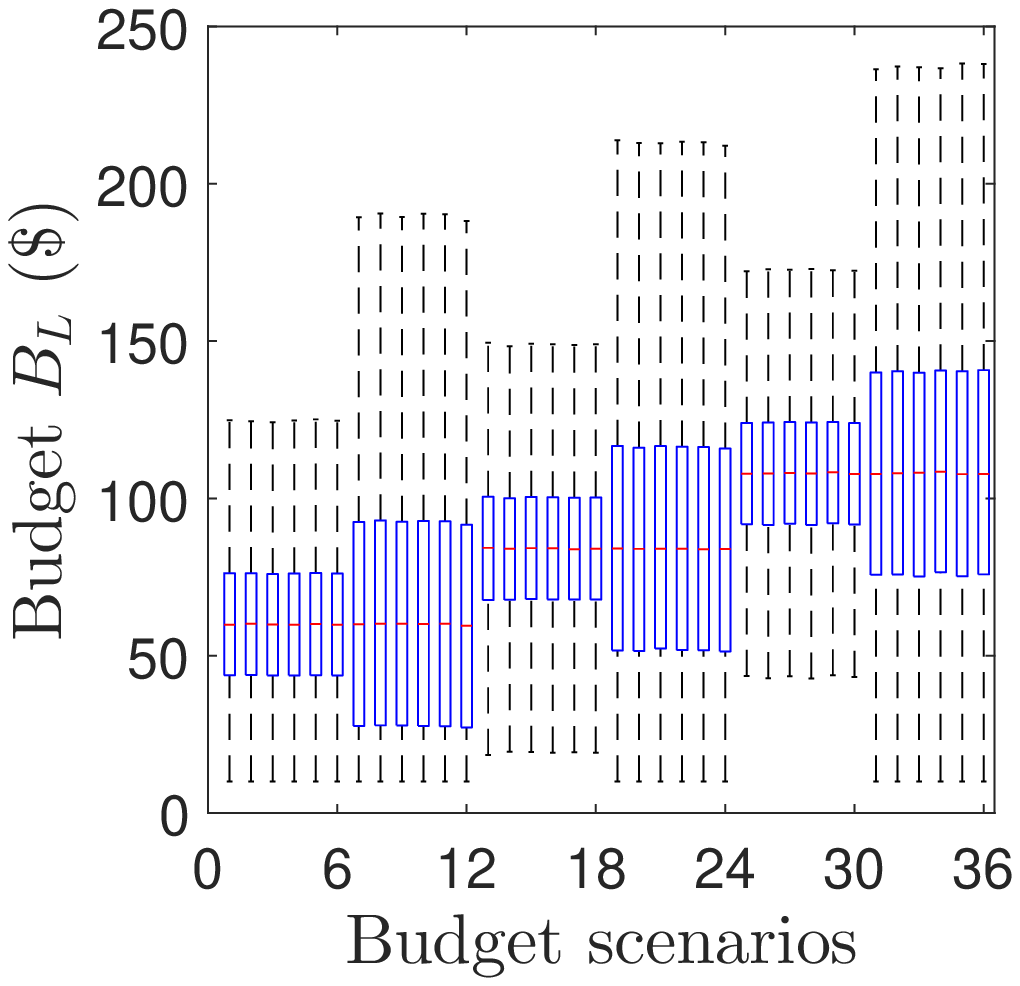} \label{fig_budget_low}}
~
\subfloat[Budget high]{\includegraphics[width=0.415\textwidth]{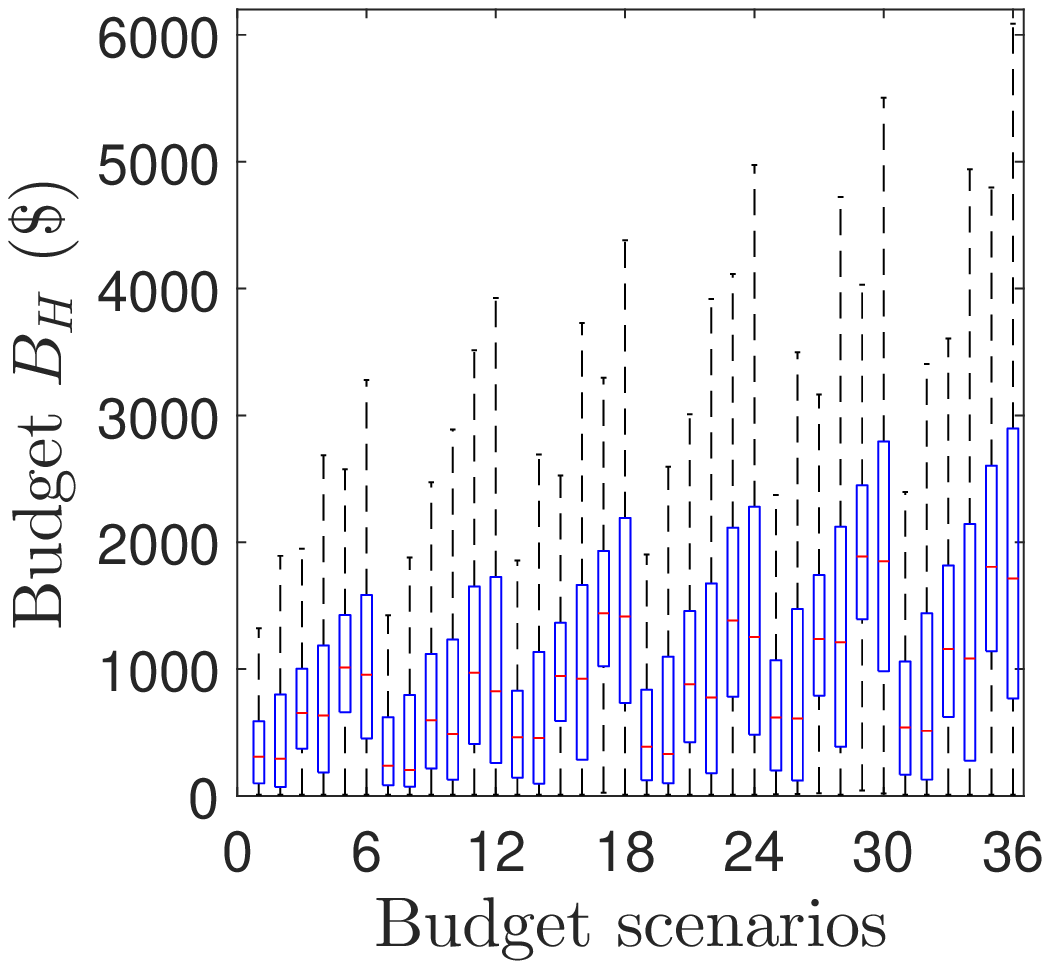} \label{fig_budget_high}}
\\
\caption{Distribution of the budgets $B_L$ and $B_H$.}
\label{fig_budget}
\end{figure}

Figs.~\ref{fig_example_PL90} and~\ref{fig_example_PL120} verify the above mentioned trends. The revenue for the low QoS scenario starts decreasing as $P_L$ increases further to $\$90$ and $\$120$, since many users cannot afford to pay these prices. The message learnt for the MNO is that, for the low QoS scenario, a high price does not lead to high revenues. Due to this, the high QoS scenario generates the highest revenue, even with very low values of $K$. Also, the maximum revenue for the high QoS scenario is admitted for a value of $K$ that decreases as $P_L$ increases. The same trends hold for the mixed QoS scenario, but with a higher value of $K$ due to fewer users with high QoS. Due to this and a steep decrease for the revenue of the high QoS scenario, the mixed QoS scenario is the most profitable when  both $P_L$ and $K$ are high.   

 \section{Revenue Analysis: General Results}
Through the detailed analysis of the previous section, we are able to compute the expected revenue of the three QoS scenarios for every possible combination of the techno-economic parameters. Though this methodology provides a \mbox{fine-grained} view for each case, we need to extract general conclusions. Indeed, for a given set of techno-economic parameters, the ultimate challenge for the MNO is to choose the prices $P_L$ and $P_H$ so that its revenue will be maximised. Therefore, we can consider this fine-grained analysis as an internal process for the MNO to compute: i)~the value of $P_L$ that maximises its revenue for the low QoS scenario, ii)~the value of $P_H$, \emph{i.e.}, parameter $K$ and $P_L$, that maximises its revenue for the high QoS scenario, and iii)~the values of $P_L$ and $P_H$ that maximise its revenue for the mixed QoS scenario. Then, the MNO can choose which QoS scenario maximises globally its revenue.
 
Though the MNO controls the technical parameters and the price, the distribution of the users' budgets as well as the users' preferences for the two QoS classes are private information. The complementary problem of how to estimate this piece of information is not addressed in this paper. However, we present a broad number of scenarios for the parameters that each user controls, so as to estimate the revenue for the three QoS scenarios under different users' behaviours.  
 
\begin{figure}[t!]
\centering
\subfloat[Revenue]{\includegraphics[width=0.4\textwidth]{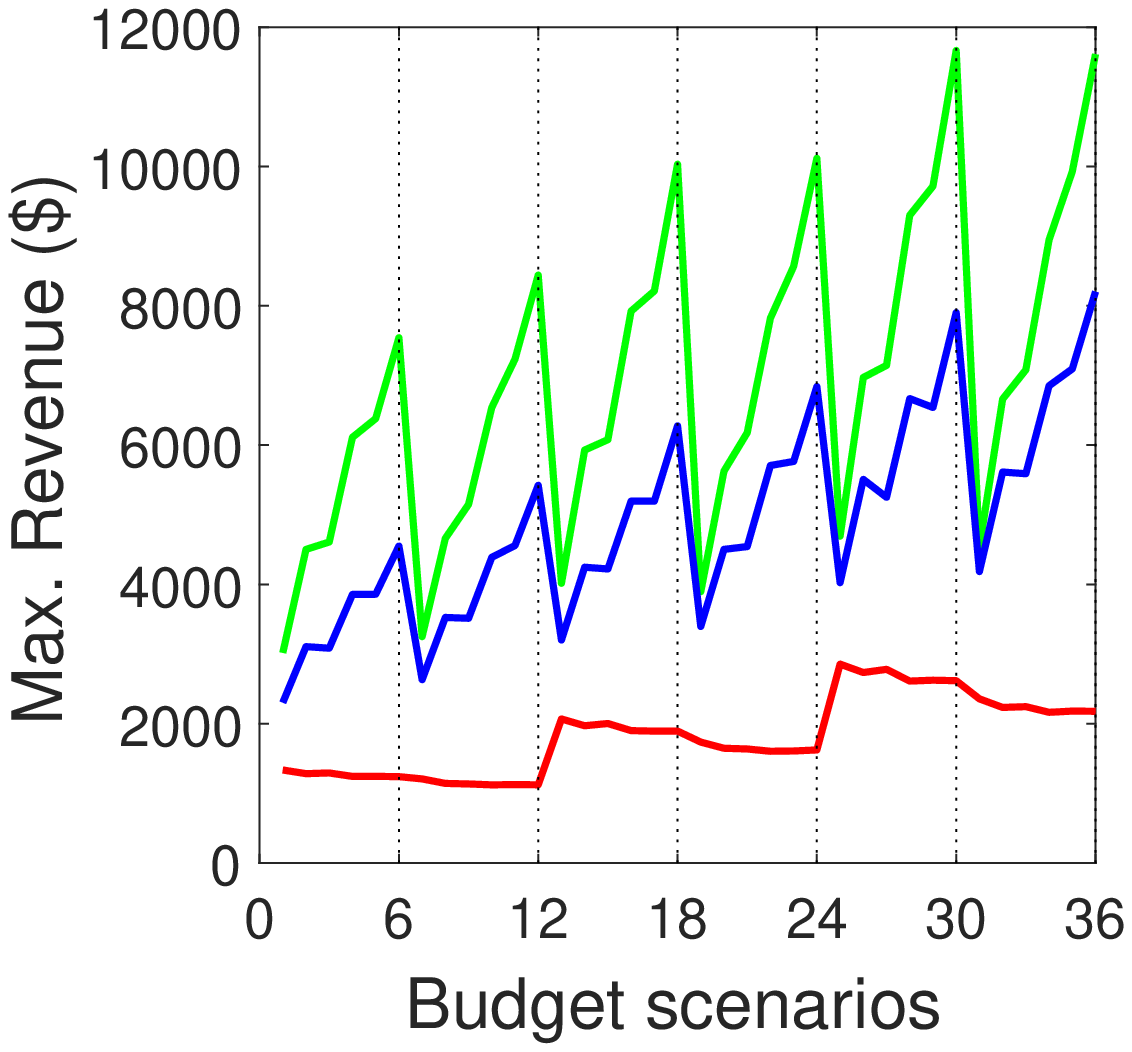} \label{fig_revenue_non_strict}}
~
\subfloat[$P_L$]{\includegraphics[width=0.38\textwidth]{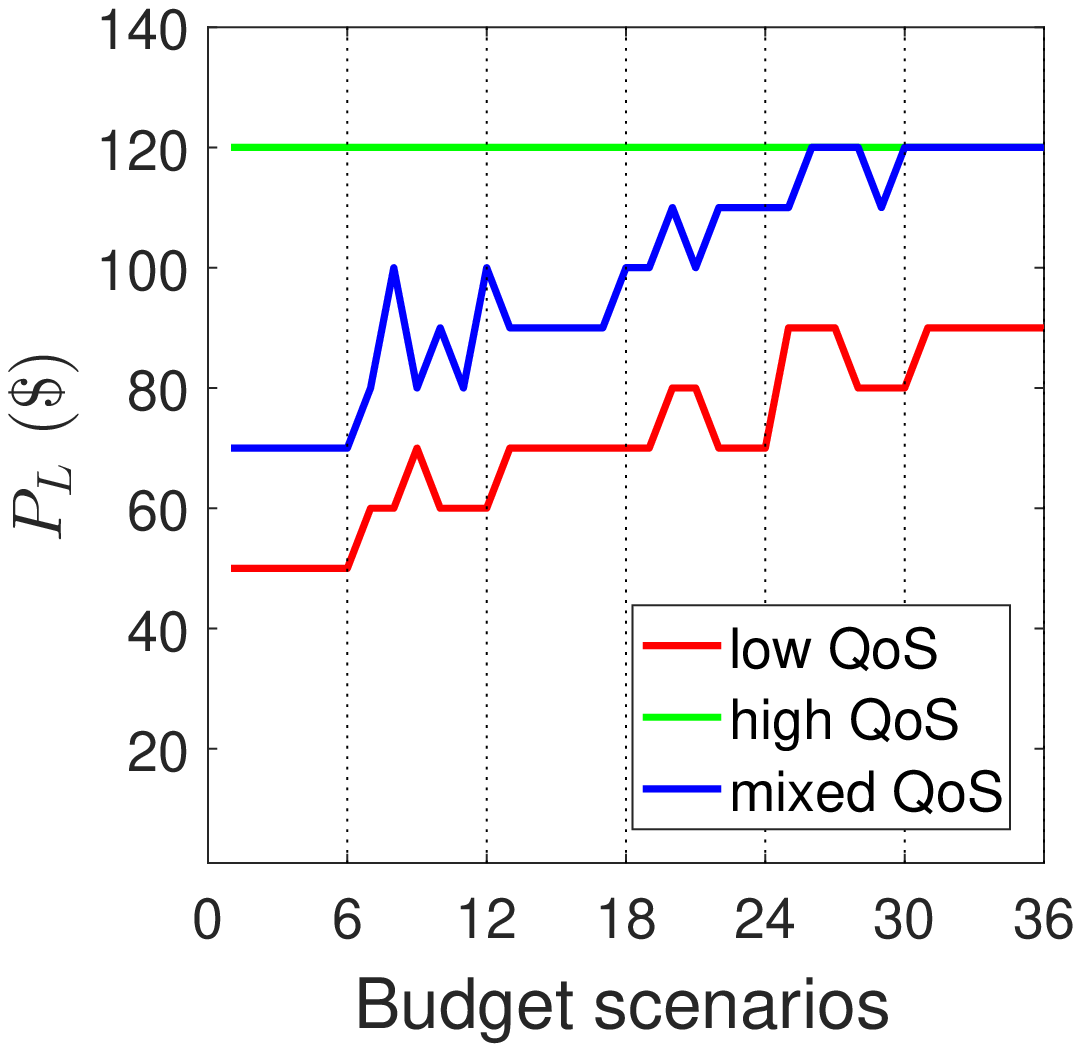} \label{fig_PL_non_strict}}
\\
\subfloat[$P_H$]{\includegraphics[width=0.4\textwidth]{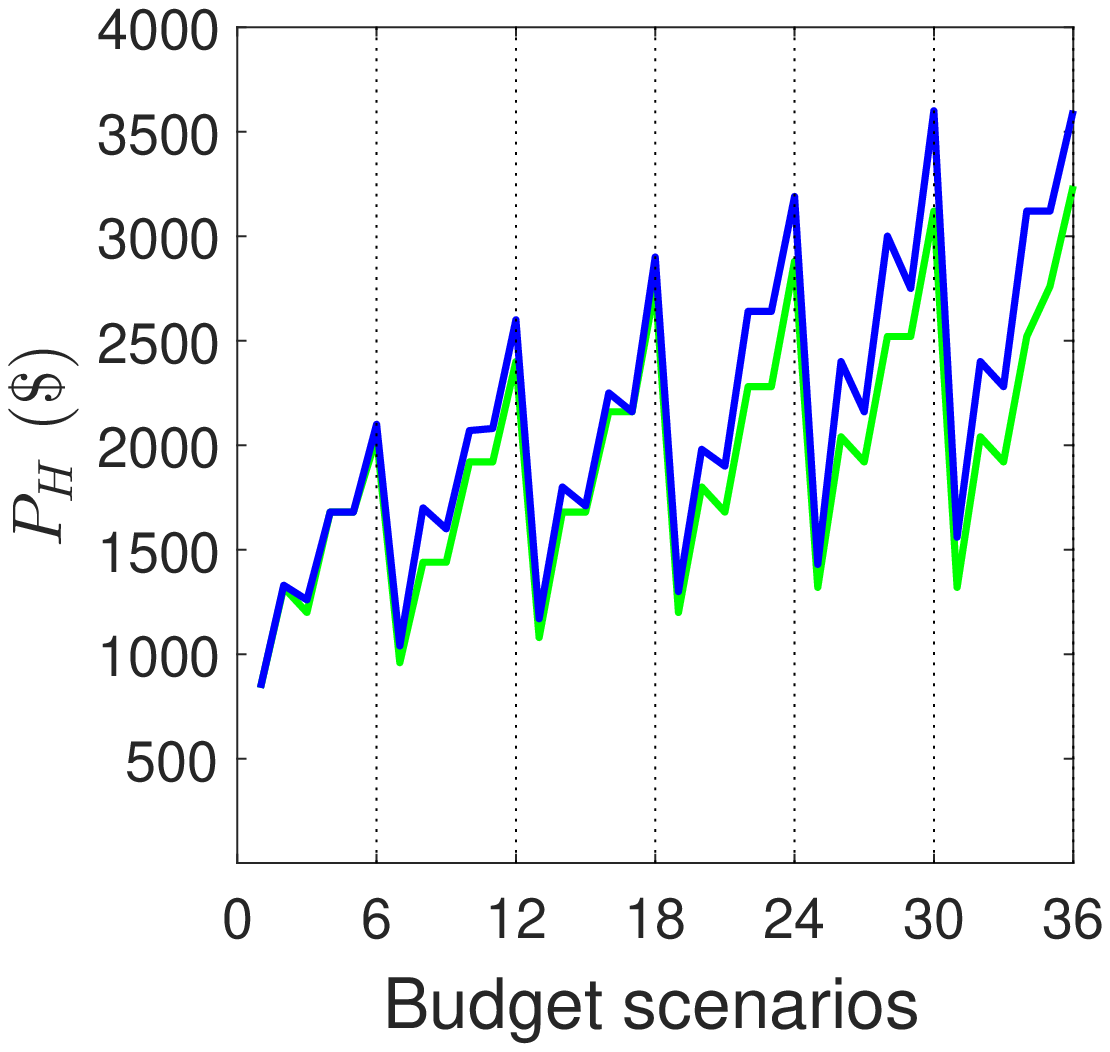} \label{fig_PH_non_strict}}
~
\subfloat[$K=\frac{P_H}{P_L}$]{\includegraphics[width=0.38\textwidth]{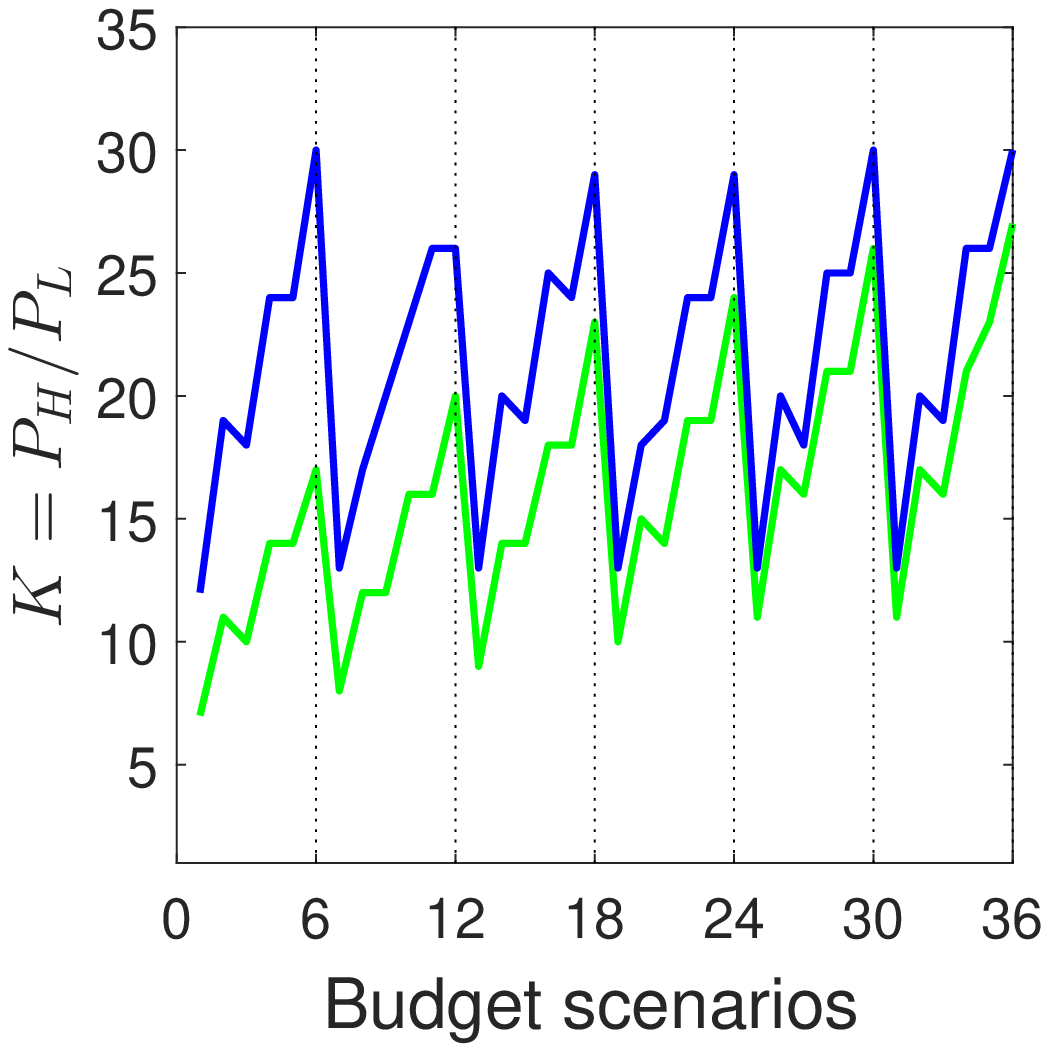} \label{fig_k_non_strict}}
\\
\caption{Max. revenue and the corresponding values for $P_L$ and $P_H$ for the three QoS scenarios. Technical parameters: $f$=3800~MHz, indoor. The choice of the QoS class is non-strict.}
\label{fig_non_strict}
\vspace{-0.4cm}
\end{figure}
 
Initially, we generalise the results of the previous section where we consider 36 budget scenarios for the distribution of the users' budgets $B_L$ and $B_H$. The number of budget scenarios arises since the 4-tuple \{$\mu_L$,  $\sigma_L$, $\mu_H$,  $\sigma_H$\} can get $3\cdot2\cdot3\cdot2=36$ possible values. Fig.~\ref{fig_budget} represents the evolution of the budget distribution. We progressively update the elements of the 4-tuple in four loops, with the following order from the outermost loop to the innermost loop: i)~$\mu_L$, ii)~$\sigma_L$, iii)~$\mu_H$, and iv)~$\sigma_H$. Due to this, as we can see from Fig.~\ref{fig_budget_low}, $\mu_L$, depicted as a red line, increases every 12 budget scenarios, remaining the same for scenarios 1-12, 13-24, and 25-36. Let us consider scenarios 1-12: due to a higher value of $\sigma_L$, scenarios 7-12 have higher upper quartiles and whiskers than scenarios 1-6. For the case of $B_H$ (Fig.~\ref{fig_budget_high}), we notice that every 6 scenarios where $\mu_L$ and $\sigma_L$ are fixed (\emph{i.e.}, scenarios 1-6, 7-12, etc.), the upper quartile increases. Moreover, the maximum upper whiskers correspond to scenarios 6, 12, etc., where $B_H$ has the highest coefficients for $\mu_H$ and $\sigma_H$. 
 
Fig.~\ref{fig_non_strict} presents the maximum revenue and the corresponding values for $P_L$ and $P_H$ for the three QoS scenarios. As in Fig.~\ref{fig_revenue_example}, we consider the \emph{non-strict} version for the choice of the QoS class and the results are obtained for the carrier frequency $f$=~3800 MHz and the indoor propagation environment. For all combinations of budgets $B_L$ and $B_H$ in Fig.~\ref{fig_revenue_non_strict}, the maximum revenue of the MNO is achieved for the high QoS scenario, followed by the mixed QoS scenario and then by the low QoS scenario. This result highlights the existence of a tussle for this market between the social welfare (\emph{i.e.}, supporting the maximum number of PMSE users) and the revenue maximisation. Focusing on the revenue from the high QoS scenario, we notice that, for budget scenarios 1-6, the maximum is for the last scenario (scenario 6) and this trend is repeated every six scenarios. The explanation is based on the previous analysis for the distribution of the budget $B_H$. The same trend holds for the mixed QoS scenario, implying that the dominant component for the mixed QoS revenue is the revenue that arises from the users with $Q_H$. Finally, for the low QoS scenario, there is a repeating trend for budget scenarios 1-12, 13-24, and 25-36. We recall from Fig.~\ref{fig_budget_low} that all budget scenarios of each of these cycles correspond to the same $\mu_L$ of the budget distribution $B_L$.  Moreover, the revenue during each cycle slightly decreases, admitting three local maxima for budget scenarios 1, 13, 25, where $\mu_H$ and $\sigma_H$ have the lowest values (see Fig.~\ref{fig_budget_high}).

\begin{figure}
\centering
\subfloat{\includegraphics[width=0.35\textwidth]{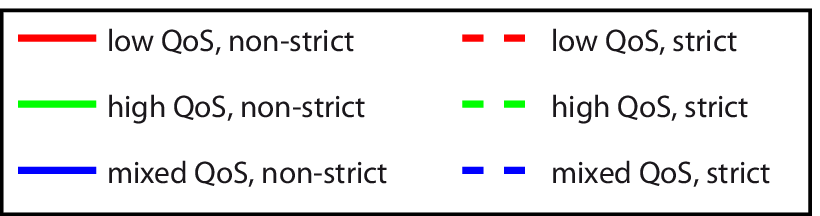}}
~
\subfloat{\includegraphics[width=0.4\textwidth]{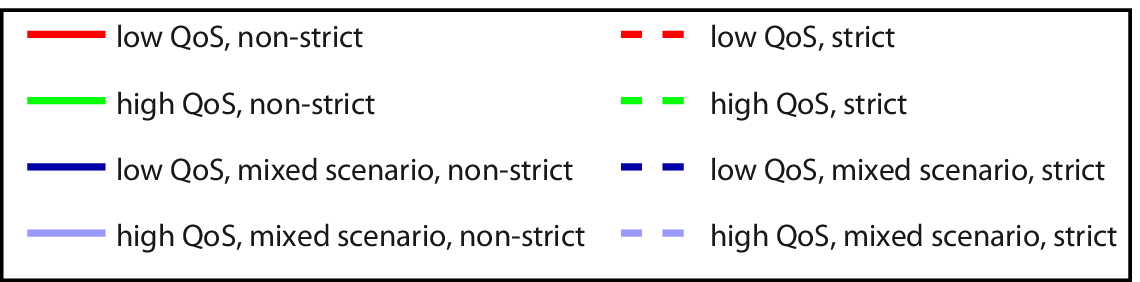}}
\\
\addtocounter{subfigure}{-2}
\subfloat[Revenue, $f$=3800~MHz, indoor]{\includegraphics[width=0.4\textwidth]{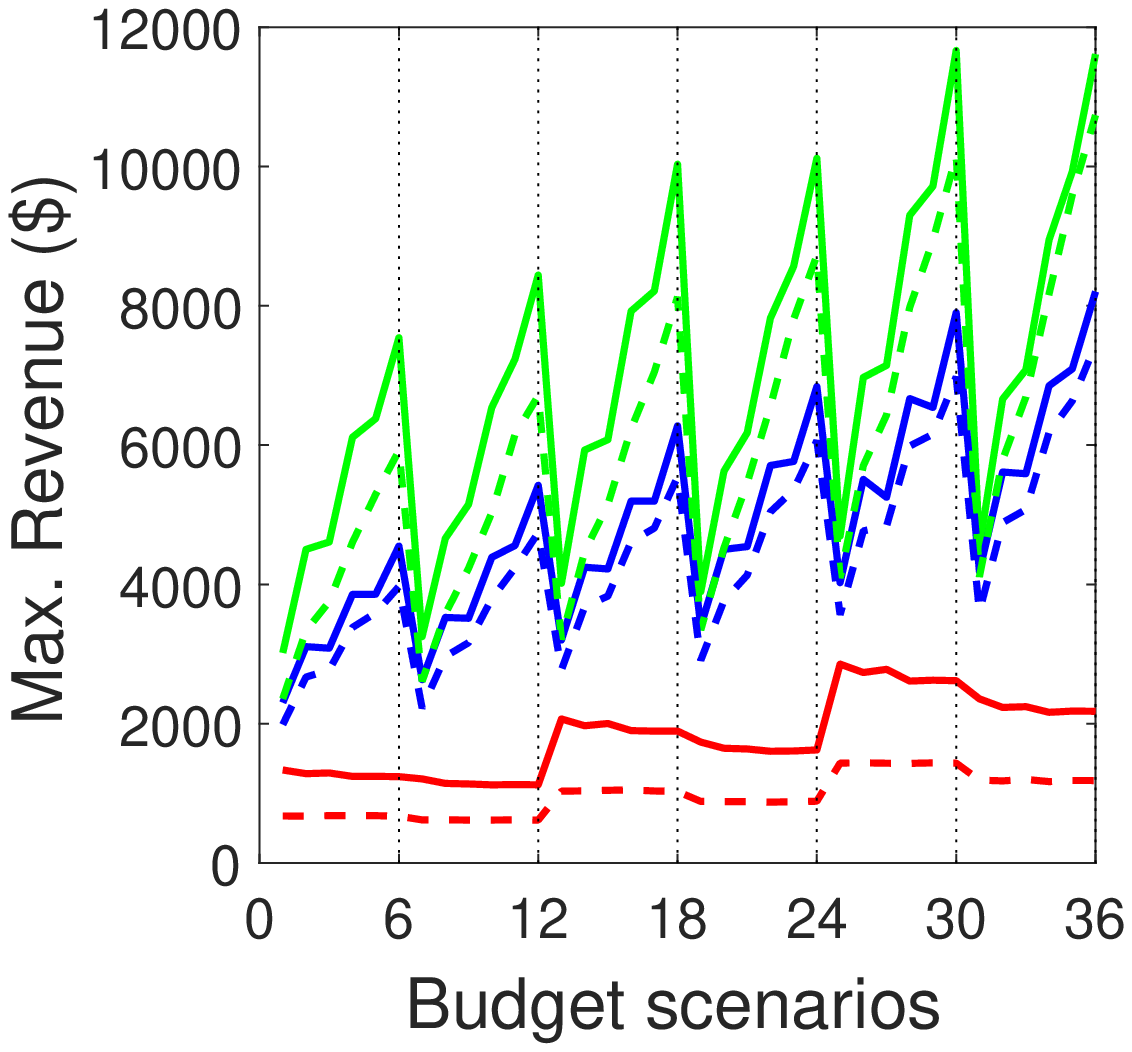} \label{fig_revenue_comparison}}
~
\subfloat[Users, $f$=3800~MHz, indoor]{\includegraphics[width=0.37\textwidth]{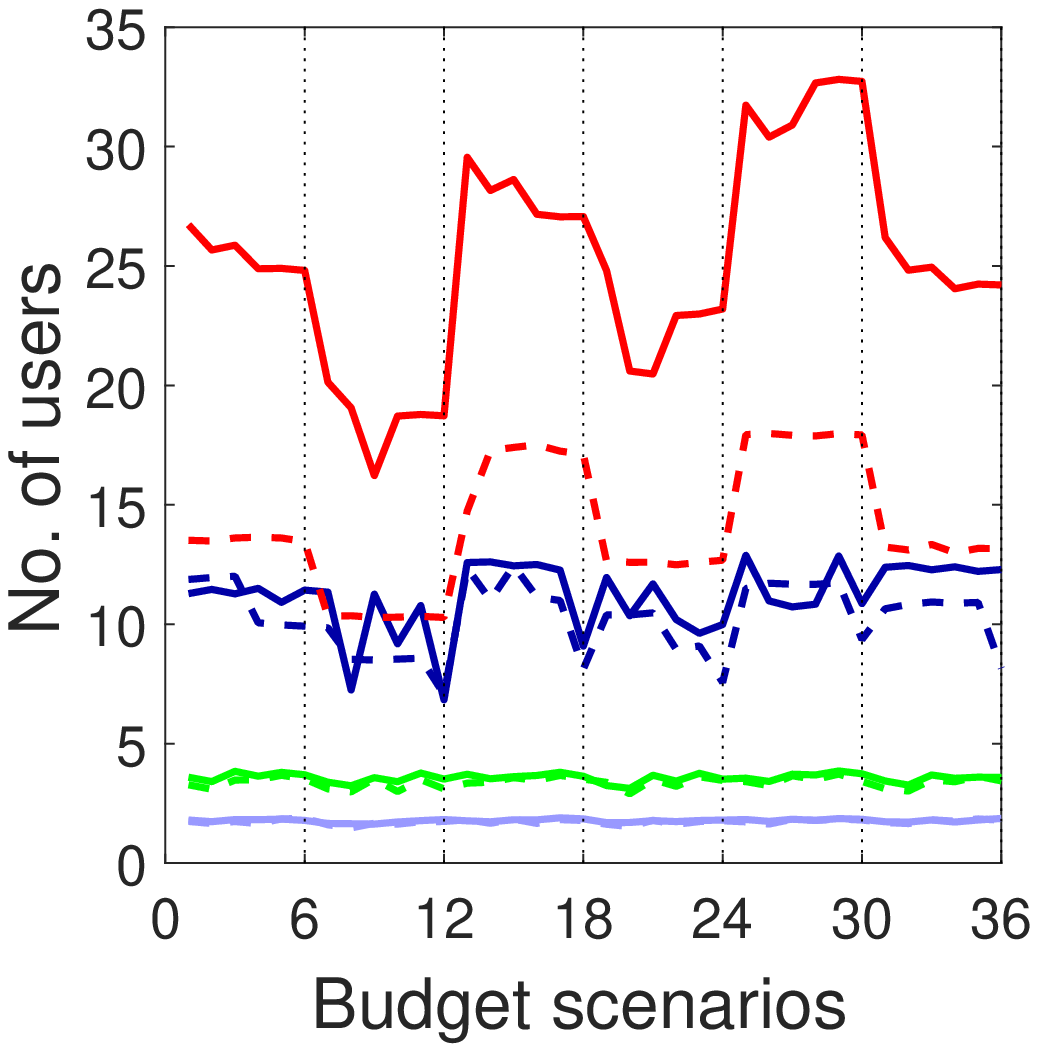} \label{fig_users_comparison}}
\\
\subfloat[Revenue, $f$=800~MHz, indoor]{\includegraphics[width=0.4\textwidth]{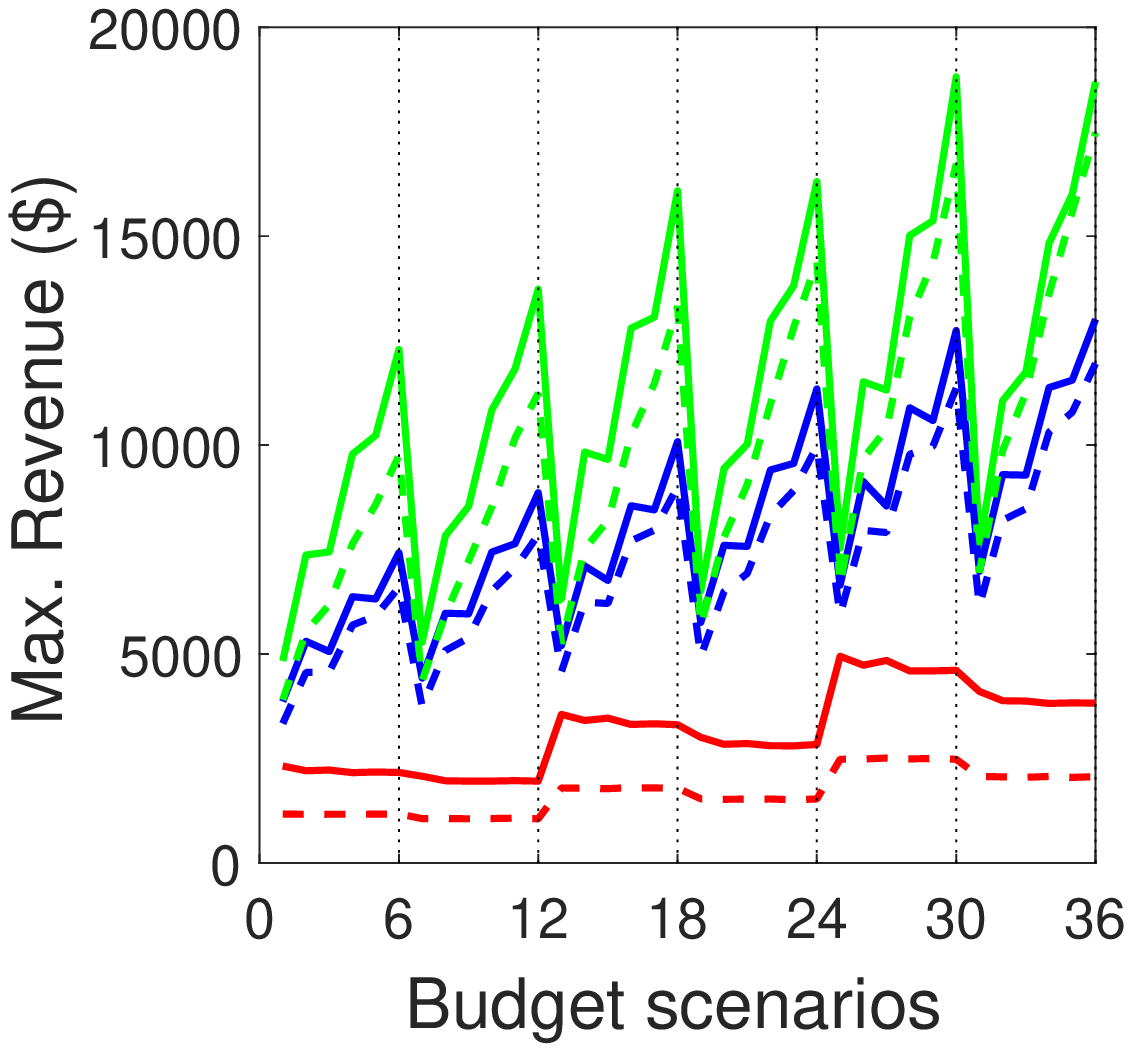} \label{fig_revenue_comparison_65}}
~
\subfloat[Users, $f$=800~MHz, indoor]{\includegraphics[width=0.37\textwidth]{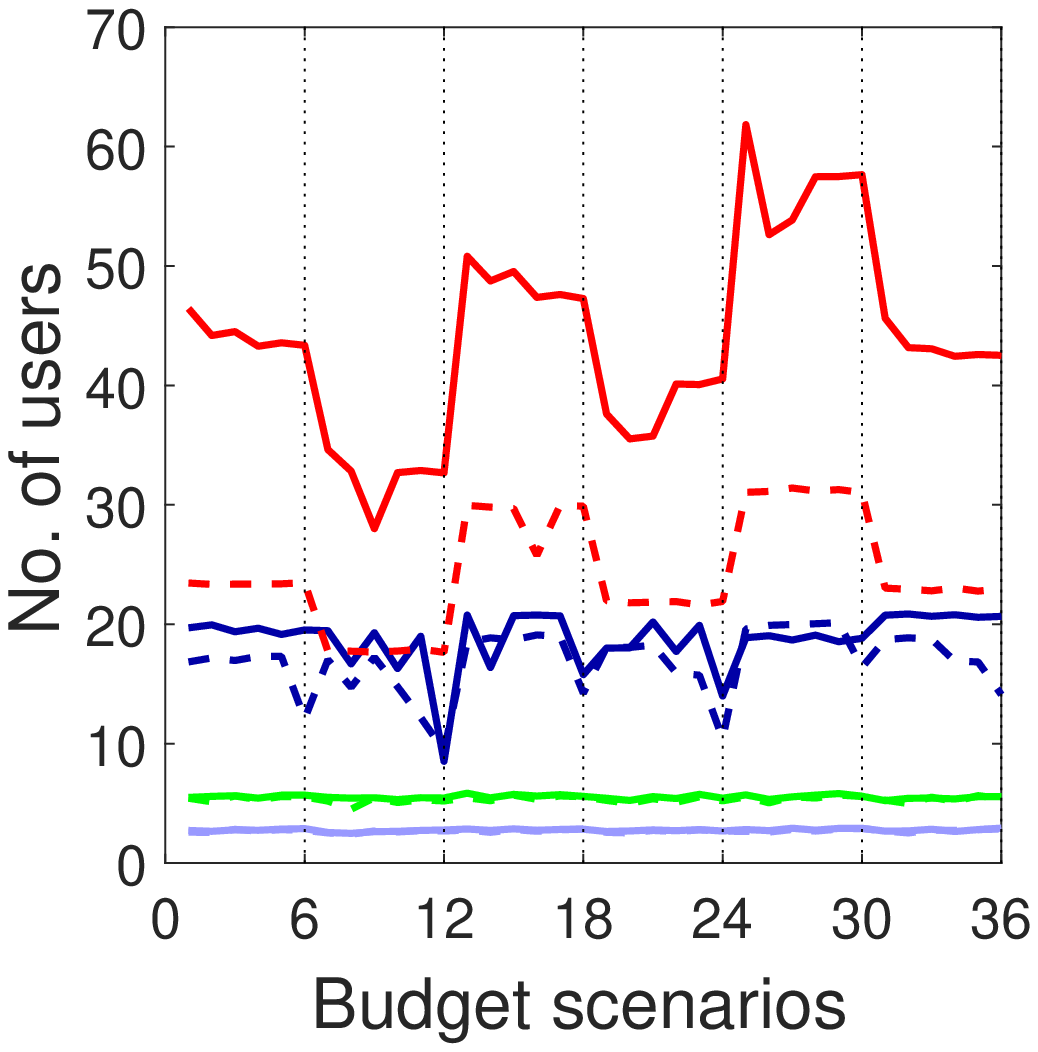} \label{fig_users_comparison_65}}
\\
\subfloat[Revenue, $f$=800~MHz, outdoor]{\includegraphics[width=0.4\textwidth]{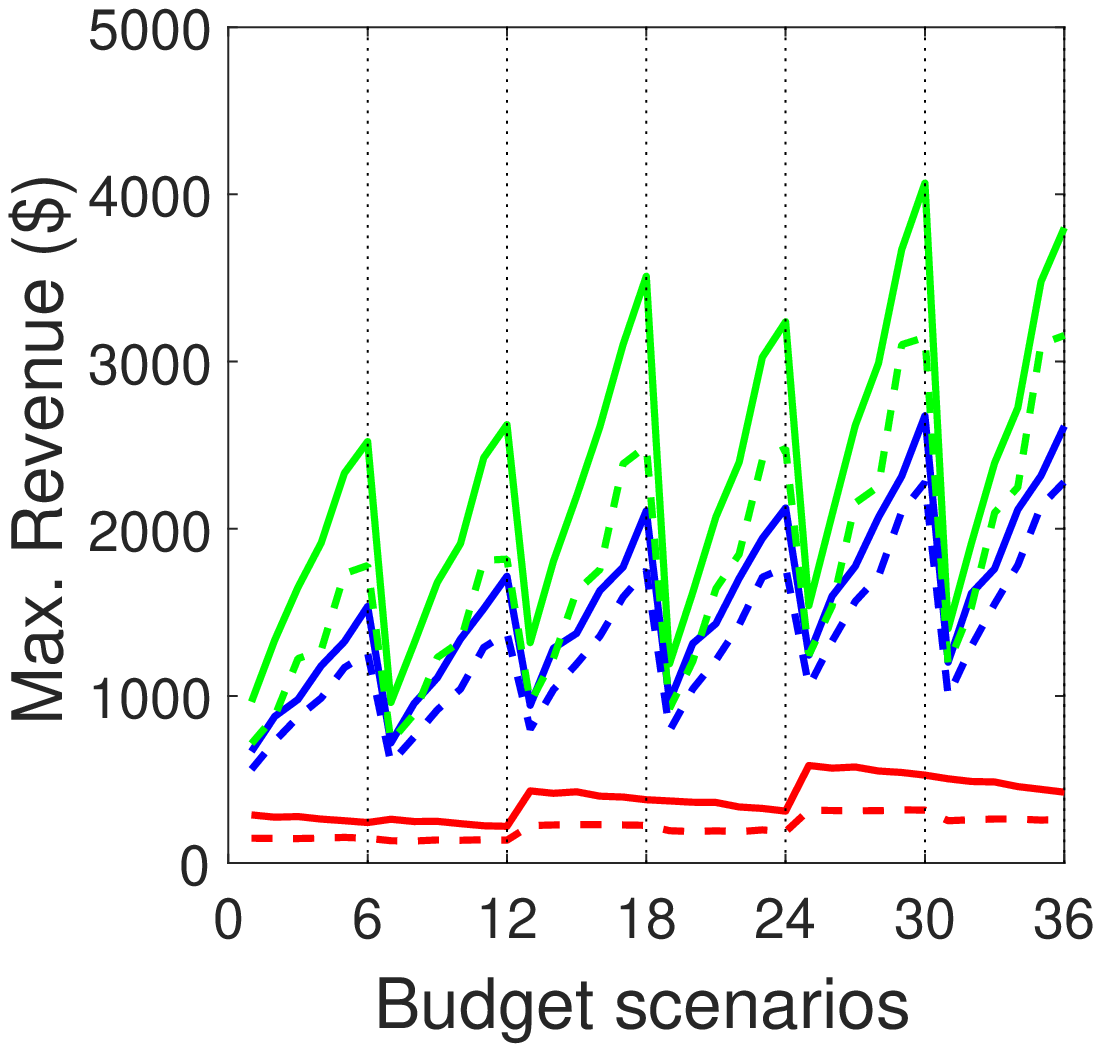} \label{fig_revenue_comparison_7}}
~
\subfloat[Users, $f$=800~MHz, outdoor]{\includegraphics[width=0.365\textwidth]{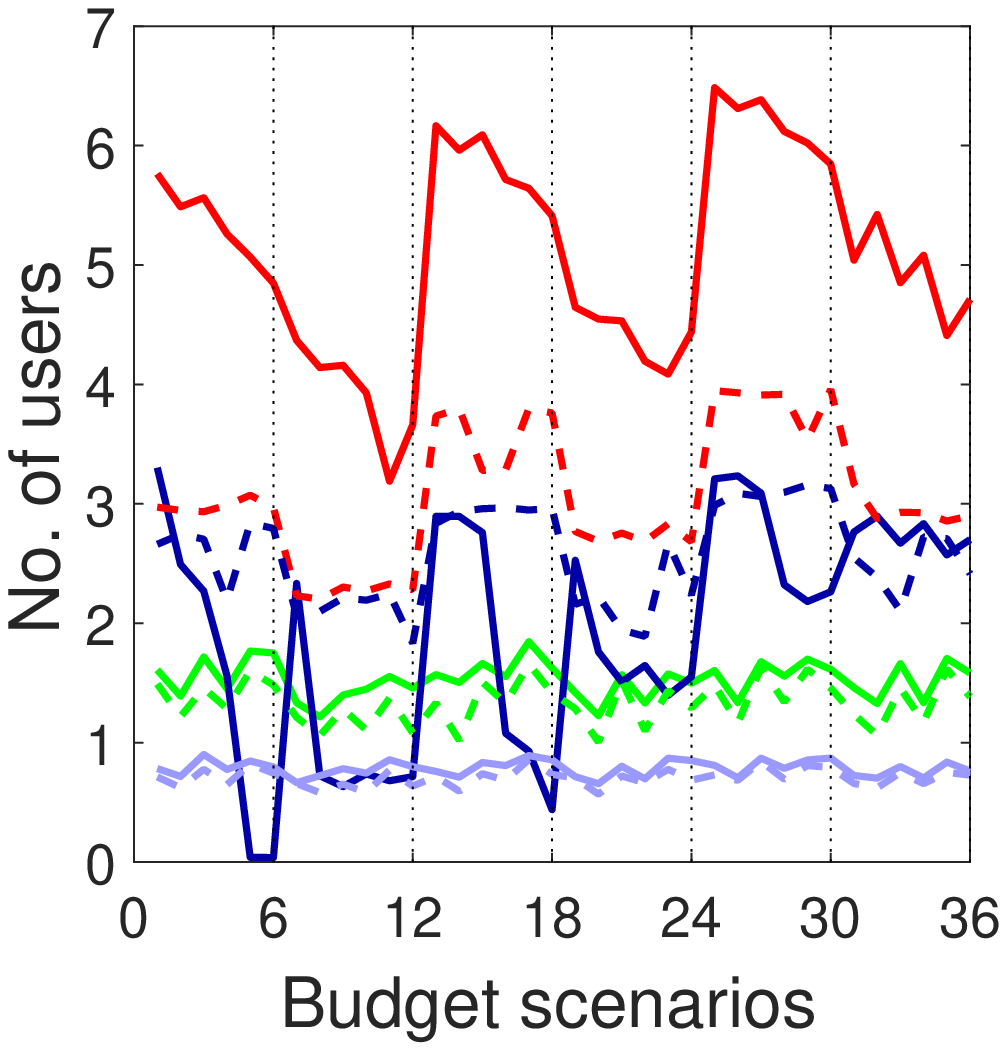} \label{fig_users_comparison_7}}\\
\caption{Comparison of the max. revenue and the corresponding number of users for the non-strict and the strict choice of the QoS class.}
\label{fig_comparison_65_7}
%\vspace{-0.4cm}
\end{figure}

Fig.~\ref{fig_PL_non_strict} presents the corresponding value of $P_L$ for which the maximum revenue for each QoS scenario is achieved. It is interesting that for the high QoS scenario, $P_L$ is always equal to $\$120$, \emph{i.e.}, the maximum that the MNO can set throughout the study. For the mixed QoS scenario, $P_L$ is higher than the corresponding price for the low QoS scenario. This is expected, since in the mixed QoS scenario, the MNO can admit at most 13 users with $Q_L$, instead of 37 users for the low QoS scenario (see Table~\ref{table_max_users}). We also notice that the evolution of $P_L$ is similar for both low and mixed QoS scenarios, with the highest values being for budget scenarios 31-36, where $\mu_L$ and $\sigma_L$ get the highest \mbox{values (see Fig.~\ref{fig_budget_low}).} 

Then, we show in Fig.~\ref{fig_PH_non_strict} the corresponding value of $P_H$. As expected, it is higher for the mixed QoS scenario where at most 2 users with $Q_H$ can be supported than for the high QoS scenario where $N_H=4$. Moreover, the curves follow the same trend with the revenue. Finally, Fig.~\ref{fig_k_non_strict}  depicts the evolution of parameter $K=\frac{P_H}{P_L}$, where the trends are similar with the trends for $P_H$. Clearly, there is room for the MNO to apply higher price differentiation for the case of the mixed QoS scenario compared to the high QoS scenario. Our analysis suggests that in budget scenarios where  $\mu_H$ and $\sigma_H$ get the highest values, the MNO has motivation to charge the mixed QoS users with $Q_H$ at the maximum level of price differentiation, \emph{i.e.}, 30 times more than the users with~$Q_L$. 

We repeat the same analysis for the \emph{strict} preference of the QoS class, where each user has a single choice for the QoS class. Fig.~\ref{fig_revenue_comparison} compares the maximum revenue for the non-strict and the strict version. The conclusion that arises is that, for all QoS scenarios and all budget scenarios, the revenue is higher for the non-strict version. This is justified due to the fact that the set of revenues for the MNO for the non-strict version is a superset of the strict version: it additionally includes the revenue that each user can bring for its second QoS preference in case it has not been admitted for its first QoS preference. We identify the factors that can justify the difference in the revenue between the non-strict and the strict version, as follows. 

The first one is that the number of PMSE users for the non-strict version can be higher than for the strict version. This is clearly the case for the low QoS scenario where, as we can see from Fig.~\ref{fig_users_comparison}, there is a significant drop in the number of users with $Q_L$ for the strict version. However, it is worth mentioning that even in the case of the non-strict version, the maximum revenue for the low QoS scenario does not coincide with the theoretical maximum of PMSE users that can be supported, which is 37. This means that either some users do not have the necessary budget $B_L$ to pay for a particular price $P_L$, or it is more profitable for the MNO to support fewer users with $Q_L$ but at a higher price. Furthermore, it is interesting to notice that, e.g., budget scenarios 1-6 correspond to a higher number of users with $Q_L$ than scenarios 7-12.  Given that these scenarios have the same mean $\mu_L$, we conclude that the standard deviation $\sigma_L$ for scenarios 1-6, which is smaller than for scenarios 7-12, is the reason for the difference in the number of users. Indeed, for the users with $Q_L$, it is more profitable for the MNO if the standard deviation $\sigma_L$ is smaller, since, for prices $P_L$ that are close to $\mu_L$, more users can afford to pay for it.

The second factor is that, in the non-strict version, the MNO may have motivation to support fewer users provided that it can charge them more.  This is the case with the mixed QoS scenario, where, for some budget parameters (budget scenarios 26-28), the MNO in the non-strict version prefers to support fewer users with $Q_L$ (dark blue solid line) than in the strict version (dark blue dashed line). 

We finally proceed with the results for the other two technical cases, \emph{i.e.}, carrier frequency $f$=800~MHz and indoor/outdoor propagation environment. We present the maximum revenue and the corresponding number of users for the three QoS scenarios in Figs.~\ref{fig_revenue_comparison_65}\textendash\ref{fig_users_comparison_7}, omitting the corresponding values of $P_L$ and $P_H$ due to space constraints. As in Fig.~\ref{fig_revenue_comparison}, the high QoS scenario generates always the highest revenue. This is a strong result independent of the technical parameters and the distribution of the budgets. Regarding the corresponding number of users, the two key conclusions that we extracted from Fig.~\ref{fig_users_comparison} still hold. First, the number of users that maximises the revenue for the low QoS scenario does not coincide with the maximum number of users (\emph{i.e.}, 65 users for indoor and 7 users for outdoor). Second, the number of users with $Q_L$ for the mixed QoS scenario is in general lower for the non-strict version compared to the strict version, since the MNO has motivation to support fewer users with $Q_L$ in order to admit more users with $Q_H$ and charge them with high values of $K$. This trend becomes clearer in Fig. \ref{fig_users_comparison_7}, where the non-strict version of the mixed QoS scenario (dark blue solid line) is almost always below the strict version of the mixed \mbox{QoS scenario} (dark blue dashed line). 

\section{Conclusions and Outlook}
The goal of this work was to unlock the potential of QoS-aware pricing for an MNO that operates under the LSA regime. The business model for the MNO was to lease spectrum to PMSE users, differentiating their prices based on whether they belong to the high or the low QoS class. We analysed three QoS scenarios: i)~all users have the same low QoS requirements, ii)~all users have the same high QoS requirements, and iii)~a mixed QoS scenario. 

From the perspective of the PMSE users, we made two contributions. First, we modelled the behaviour of the users regarding how they choose between the two QoS classes, quantifying the importance that each user gives to the QoS class versus the price that it has to pay. Second, we modelled the distribution of the budget of the users for the two QoS classes. The added value of these models is that we were able to perform a fine-grained analysis, predicting the distribution of the users between the two QoS classes for each possible combination of considered prices.  

From the perspective of the MNO, the challenge was to choose the prices $P_L$ and $P_H$ so as to compute the maximum revenue that can be achieved for each QoS scenario. Our analysis revealed a consistent result that holds independent of i) the distribution of the budgets, ii) the way that the users choose between the QoS classes, and iii) the values of the technical parameters. The MNO can always tune the prices so that the maximum revenue for the high QoS scenario is the highest, followed by the mixed QoS scenario and finally by the low QoS scenario. This result highlights the potential of QoS-aware pricing for the MNO. For the high and mixed QoS scenarios where QoS price differentiation can be applied, the MNO can consistently generate higher revenue than for the low QoS scenario. This is also interesting from a regulatory point of view, since the MNO has motivation to support few users charging them at a higher price instead of supporting more users at a lower price. Therefore, we identified a constant tussle in the LSA market, where the goal of the MNO (\emph{i.e.}, revenue maximisation) is not aligned with the goal of the market regulator (\emph{i.e.}, social welfare maximisation). 

Through the analysis of the revenues for the different budget scenarios, we identified the impact of the budget parameters on the revenue of the QoS scenarios. The revenue for the high and mixed QoS scenarios admits local maxima when both the mean and the standard deviation of the budget distribution $B_H$ are high (budget scenarios 6, 12, etc.). On the other hand, the revenue for the low QoS scenario admits local maxima when the mean of the budget distribution $B_L$ and both parameters of the budget distribution $B_H$ are small (budget scenarios 1, 13, 25). These trends hold for any values of the technical parameters. We argue that they are useful in particular for an MNO who evaluates the business opportunities in different markets before entering into them since they provide insights for which markets have the potential to be more profitable. 

Finally, we conclude with two key messages extracted from our study for the mixed QoS scenario. First, there is higher room for price differentiation for the mixed QoS scenario, since fewer users with $Q_H$ can be admitted compared to the high QoS scenario. Second, for the non-strict version of the choice of the QoS class, the MNO usually prefers to sacrifice some of the users with $Q_L$ in order to support more users with $Q_H$ and charge them more. Both conclusions reinforce the message learnt, \emph{i.e.}, that the application of QoS-aware pricing unlocks significant revenue opportunities. 

As future work, it is interesting to extend this study by introducing an additional (intermediate) QoS class and evaluate the robustness of the results. This also requires a modification for the way that the users choose among the three QoS classes. Another interesting direction is to consider an oligopoly market with two or three MNOs, analysing the churn of the users and the evolution of the revenue as the MNOs update their pricing policies. 

\section*{Acknowledgment}
The authors would like to thank Shaham Shabani who conducted the simulations in ns-3 for estimating the maximum number of PMSE users for all QoS scenarios.   

%\vspace{-0.3cm}  
\bibliographystyle{IEEEtran}
\bibliography{IEEEabrv,references}
\end{document}